\documentclass[twocolumn]{aastex6}

\collaborationName{}
\begin{document}

\title{\emph{K2} Discovers a Busy Bee: An Unusual Transiting Neptune Found in the Beehive Cluster} 

\author{
Christian Obermeier\altaffilmark{1,2,3},
Thomas Henning\altaffilmark{1},
Joshua E. Schlieder\altaffilmark{4,5,6}, 
Ian J.\ M.\ Crossfield\altaffilmark{7,8}, 
Erik A.\ Petigura\altaffilmark{9,10},
Andrew W.\ Howard\altaffilmark{11},
Evan Sinukoff\altaffilmark{11,12},
Howard Isaacson\altaffilmark{13},
David R.\ Ciardi\altaffilmark{6},
Trevor J.\ David\altaffilmark{14},
Lynne A.\ Hillenbrand\altaffilmark{14},
Charles A.\ Beichman\altaffilmark{13},
Steve B.\ Howell,\altaffilmark{4},
Elliott Horch\altaffilmark{15},
Mark Everett\altaffilmark{16},
Lea Hirsch\altaffilmark{13},
Johanna Teske\altaffilmark{17,18},
Jessie L. Christiansen\altaffilmark{6},
S\'ebastien L\'epine\altaffilmark{19},
Kimberly M.\ Aller\altaffilmark{11},
Michael C. Liu\altaffilmark{11},
Roberto P. Saglia\altaffilmark{2},
John Livingston\altaffilmark{20},
}

\and

\author{Matthias Kluge\altaffilmark{2,3}}

\altaffiltext{1}{Max Planck Institut f\"ur Astronomie, Heidelberg, Germany}
\altaffiltext{2}{Max-Planck-Institute for Extraterrestrial Physics, Garching, Germany}
\altaffiltext{3}{University Observatory Munich (USM), Ludwig-Maximilians-Universit\"at, Munich, Germany}
\altaffiltext{4}{NASA Ames Research Center, Moffett Field, CA, USA}
\altaffiltext{5}{NASA Postdoctoral Program Fellow}
\altaffiltext{6}{NASA Exoplanet Science Institute, California Institute of Technology, Pasadena, CA, USA}
\altaffiltext{7}{Lunar \& Planetary Laboratory, University of Arizona, 1629 E. University Blvd., Tucson, AZ, USA}
\altaffiltext{8}{NASA Sagan Fellow}
\altaffiltext{9}{Geological and Planetary Sciences, California Institute of Technology, Pasadena, CA, USA}
\altaffiltext{10}{Hubble Fellow}
\altaffiltext{11}{Institute for Astronomy, University of Hawai`i at M\={a}noa, Honolulu, HI, USA}
\altaffiltext{12}{NSERC Postgraduate Research Fellow}
\altaffiltext{13}{Astronomy Department, University of California, Berkeley, CA, USA}
\altaffiltext{14}{Department of Astronomy, California Institute of Technology, Pasadena, CA, USA}
\altaffiltext{15}{Department of Physics, Southern Connecticut State University, New Haven, CT, USA}
\altaffiltext{16}{National Optical Astronomy Observatory, Tucson, AZ, USA}
\altaffiltext{17}{Carnegie Department of Terrestrial Magnetism, Washington, DC, USA}
\altaffiltext{18}{Carnegie Origins Postdoctoral Fellow, jointly appointed by Carnegie DTM and Carnegie Observations}
\altaffiltext{19}{Department of Physics and Astronomy, Georgia State University, GA, USA}
\altaffiltext{20}{Department of Astronomy, Graduate School of Science, The University of Tokyo, 7-3-1 Bunkyo-ku, Tokyo 113-0033, Japan}

\begin{abstract}
Open clusters have been the focus of several exoplanet surveys but only a few planets have so far been discovered. The \emph{Kepler} spacecraft revealed an abundance of small planets around small, cool stars, therefore, such cluster members are prime targets for exoplanet transit searches. \emph{Kepler}'s new mission, \emph{K2}, is targeting several open clusters and star-forming regions around the ecliptic to search for transiting planets around their low-mass constituents.  Here, we report the discovery of the first transiting planet in the intermediate-age (800\,Myr) Beehive cluster (Praesepe). K2-95 is a faint ($\mathrm{Kp = 15.5\,mag}$) $\mathrm{M3.0\pm0.5}$ dwarf from \emph{K2}'s Campaign 5 with an effective temperature of $\mathrm{3471 \pm 124\,K}$, approximately solar metallicity and a radius of $\mathrm{0.402 \pm 0.050 \,R_\odot}$. We detected a transiting planet with a radius of $\mathrm{3.47^{+0.78}_{-0.53} \, R_\oplus}$ and an orbital period of 10.134 days.
We combined photometry, medium/high-resolution spectroscopy, adaptive optics/speckle imaging and archival survey images to rule out any false positive detection scenarios, validate the planet, and further characterize the system. The planet's radius is very unusual as M-dwarf field stars rarely have Neptune-sized transiting planets. The comparatively large radius of K2-95b is consistent with the other recently discovered cluster planets K2-25b (Hyades) and K2-33b (Upper Scorpius), indicating systematic differences in their evolutionary states or formation. These discoveries from \emph{K2} provide a snapshot of planet formation and evolution in cluster environments and thus make excellent laboratories to test differences between field-star and cluster planet populations. 

\end{abstract}


\section{INTRODUCTION}

Exoplanet science is still a young field, but what stands out is the strong diversity in the properties of both detected planets and their host stars. Already a short time after the first 
transiting planet was detected by \citet{2000ApJ...529L..45C,2000ApJ...529L..41H}, surveys were started with a focus on open clusters for a variety of reasons. The higher density of stars gives surveys access to more stars for a given field of view. Age, distance and metallicity of the member stars are well determined, yielding more precise estimates for the planetary and stellar parameters. Furthermore, most observed field stars are relatively old ($\ge$ 1\,Gyr) while many currently targeted clusters present a younger sample 
(10-800\,Myr). In addition, planet formation in stellar clusters may well be very different due to stronger and more frequent gravitational interactions between the stars. Planets in younger clusters may also be undergoing thermal evolution, radial contraction, or receiving high irradiation from their active host stars. 
Therefore, open clusters are an 
excellent laboratory to test planet formation and evolution models. 
Initial transit surveys that focussed on 47 Tuc \citep{2000ApJ...545L..47G,2005ApJ...620.1043W}, NGC 2301 \citep{2005PASP..117.1187H} and NGC 7789 \citep{2006MNRAS.367.1677B}, 
found no evidence for transiting planets. Since then, fourteen planets have been discovered in open clusters, namely in NGC 6811 \citep{2013Natur.499...55M}, NGC 2423 \citep{2007A&A...472..657L}, M67 \citep{2014AA...561L...9B,2016arXiv160605247B}, the Beehive (Praesepe) \citep{2012ApJ...756L..33Q,2016A&A...588A.118M}, the Hyades \citep{2007ApJ...661..527S,2014ApJ...787...27Q, 2015arXiv151200483M, 2016AJ....151..112D} and Upper Scorpius \citep{2016Nature...accepted, 2016arXiv160406165M}. All planets in M67, the planet in NGC 2423, one planet in the Hyades and the Praesepe 
planets were detected with the radial velocity (RV) method.  All planets in NGC 6811, one planet in the Hyades and the planet in Upper Scorpius were discovered with the transit method. %All detections were of large, Neptune- to Jupiter-size planets.
All detections were of planets that likely harbor significant gaseous envelopes. Additionally, a $\sim$2 Myr old hot Jupiter located in the Taurus-Auriga star forming region was detected via the RV method \citep{2016Natur.534..662D}. 

All transiting cluster planets were detected with the \textit{Kepler} space telescope.
After the failure of two of its four reaction wheels, the original mission of \emph{Kepler} ended and was redirected for the "second light" survey \emph{K2} 
\citep{2014PASP..126..398H}. Instead of continuously observing the same area over years, the \emph{K2} mission switches fields every three months, stabilized by the two 
remaining reaction wheels and solar photon pressure for the third axis (roll angle). However, the telescope still drifts slowly and has to be corrected by firing the thrusters every 6 hours. 
Photometric precision is therefore slightly lower than during the \emph{Kepler} mission but, as will be described in the following section, can be corrected very well.

The Beehive cluster (M44), also called Praesepe, is an open cluster targeted by \emph{K2} in Campaign 5. It is nearby \citep[$\mathrm{d = 183 \pm 8\,pc}$,][]{2009A&A...497..209V,2011JAVSO..39..219M} and of intermediate age. Past estimates placed the age of Praesepe at around 600\,Myr \citep{2008A&A...483..891F} but new estimates that take into account the effects of rotation in its high-mass members suggest an age as old 800\,Myr \citep{2015ApJ...807...58B}. Furthermore, the kinematics \citep{2002A&A...381..446M}, metallicity \citep{2006MNRAS.369..383D} and age \citep{2015ApJ...807...58B} of Praesepe are very similar to the Hyades cluster. The age of Hyades was also redetermined to 800\,Myr \citep{2015ApJ...804..146D,2015ApJ...807...24B} and it is now assumed that both clusters may share the same origin.

Since the transit signal gets stronger with decreasing stellar radius, M dwarfs are promising targets for the detection of small planets in an open cluster. \citet{2015ApJ...807...45D} estimate an abundance of rocky and small sub-Neptunian planets around those stars with periods shorter than 200 days with an average of $2.5 \pm 0.2$ planets per star with radii between $1-4\,R_\oplus$.
Here, we present the discovery and validation of a transiting Neptune-sized planet in the Praesepe cluster detected in \emph{K2} Campaign 5 in orbit around the low-mass star K2-95. In  \S2 we 
describe the layout of our photometric and spectroscopic follow-up and detail the subsequent results in \S3. We validate the candidate as a planet in \S4, discuss the impact of our findings in the context of exoplanets in clusters and the field in \S5, and provide concluding remarks in \S6.

\section{OBSERVATIONS}

\subsection{K2 target selection and photometry}

We identified the star K2-95 as a potential M dwarf target and high probability member of the Praesepe cluster for our \emph{K2} Campaign 5 proposal (GO5006 - PI Schlieder). 
Other groups also proposed this star as a potential \emph{K2} target (GO5011 - PI Beichman, GO5048 - PI Guzik, GO5095 - PI Agueros, GO5097 - PI Johnson).

K2-95 was observed during \emph{K2} Campaign 5 with nearly continuous photometry from 2015 Apr 27 to 2015 Jul 10. We extracted the photometry from the pixel data which 
we downloaded from the MAST\footnote{The Mikulski Archive for Space Telescopes.}. 

\begin{figure*}[hbt]
\centering
\includegraphics[width=1\linewidth]{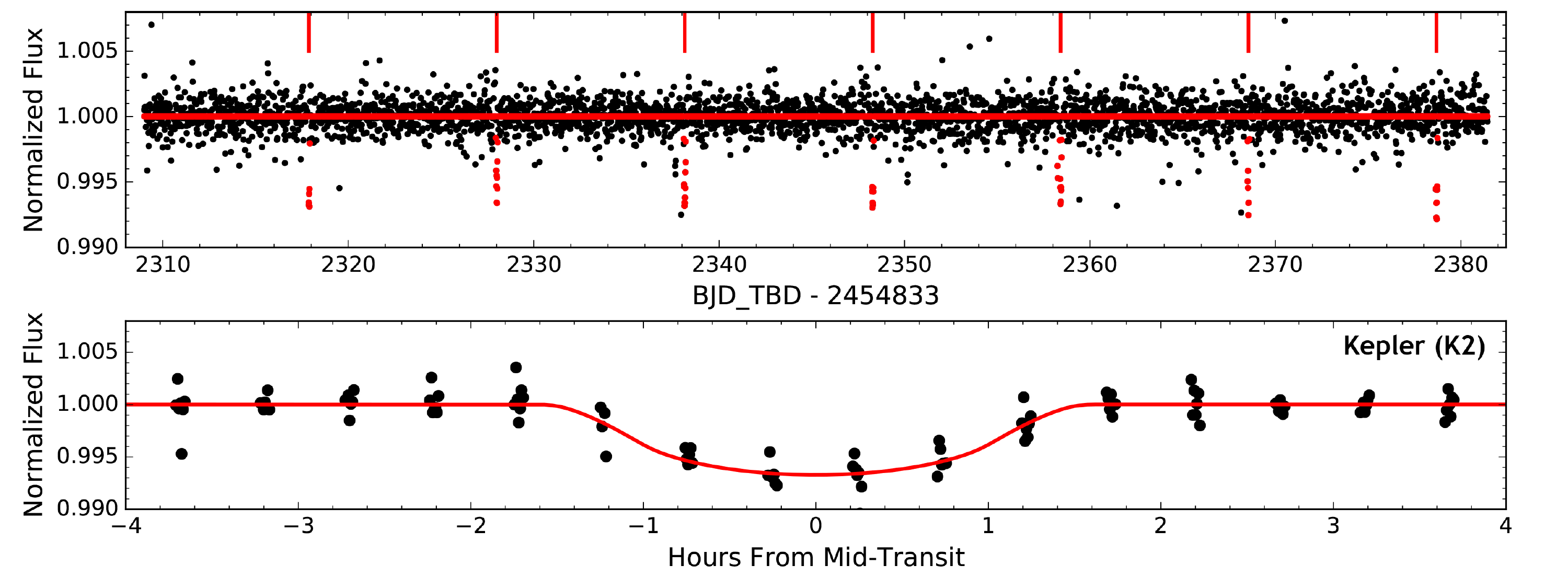}
\caption{Top: Calibrated and normalized K2 photometry for K2-95. The upper red lines indicate the detected transits with the corresponding points also marked in red. Bottom: Period-folded light curve with the best-fitting transit model overlaid as a red line.}
\label{fig:folded-lc}
\end{figure*}

Our photometric extraction pipeline is described in more detail in \citet{2015ApJ...811..102P} and \citet{2015ApJ...804...10C}. 
During \emph{K2} operations, the telescope is torqued by solar radiation pressure which causes it to slowly roll around the boresight. This motion causes stars to drift across the CCD by about 1 pixel every 6 hours.  As stars are sampled by different pixels, intra-pixel sensitivity 
and flat-fielding variations cause the apparent brightness of the star to change. Thruster fires to 
correct for this drift affect the pointing and therefore pixel position greatly, giving the overall photometry a saw-tooth shape. 
We solve for the roll angle between each frame and an arbitrary reference frame and model the time- and roll-dependent brightness variations using a Gaussian process. Further, we 
adjust the size of our square extraction aperture to minimize the residual noise in the corrected light curve. This balances two competing effects: larger apertures yield smaller 
systematic errors while smaller apertures include less background noise. %The square extraction aperture (r = 1 pixel $\approx 4''$) is shown in Figure \ref{fig:AI} on the left panel. 
Our final square extraction aperture is r = 1 pixel $\approx 4''$.
The resulting, de-trended light curve exhibited slow, periodic, $\sim$1\% modulations with a period of about 24 days. We attribute this modulation to spots on the rotating stellar surface. The timescale of this variation is long compared to other M dwarfs in Praesepe and places K2-95 among the slowest rotators in the cluster (see also section \ref{sec:kin}).
This variation is fitted and removed to produce the final light curve which is shown in the top panel of Figure \ref{fig:folded-lc}.

We searched through the optimized light curve with the TERRA algorithm which is described in more detail by \citet{2013ApJ...770...69P}. In short, it searches for periodic box-shaped photometric dimmings and fits them with a model from \citet{2002ApJ...580L.171M}. Using TERRA, we detected a transit signal in the K2-95 light curve with a period of $
\mathrm{P=10.132\,d}$ and a signal-to-noise ratio (SNR) of 23.97. The phase-folded light curve is shown in the bottom panel of Figure \ref{fig:folded-lc}, centered around the transit event. We subtracted the best-fitting model transit and iterated the TERRA algorithm to search for other transits but did not detect any secondary signals. 
%\textbf{We require three detected periods for a successful detection which limits the highest possible periods to 25 days with an overall observing time of 75 days. However, after removing 
%It is not possible to identify periods longer than 25 days since the algorithm requires at least three visible transits for a detection and the overall observing time was 75 days. However, 
Visual inspection also did not reveal any additional transit features.

\subsection{Photometric follow-up}

\begin{figure*}[hbt]
\centering
\includegraphics[width=1\linewidth]{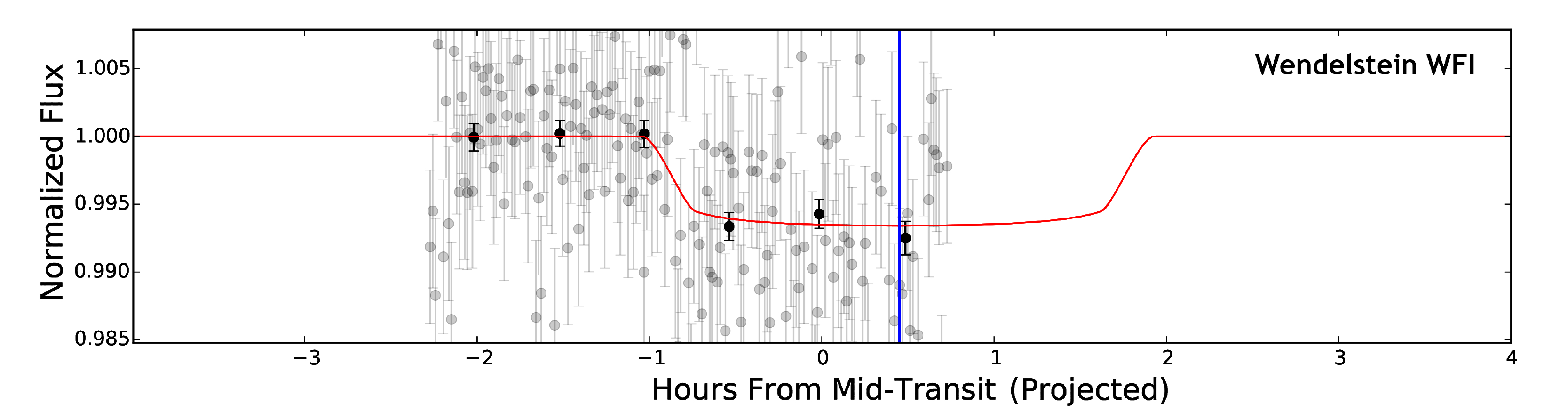}
\caption{Normalized photometry in the i'-band for K2-95, recorded with the Wendelstein WFI. We overlaid the best-fitting transit model from the \emph{K2} data, adapted with appropriate quadratic limb darkening parameters for the i'-band. The binned points (black) agree very well with the model (red line), however, the transit was shifted by about 27\,min (new center indicated by the blue line) which indicates an error in the initial period estimate within the fitting uncertainties. The original points (light grey) are shown in the background.}
\label{fig:WST-transit}
\end{figure*}

We observed K2-95 with the 2.0\,m Fraunhofer Telescope Wendelstein (FTW) \citep{2014SPIE.9145E..2DH}, using the Wide Field Imager (WFI) \citep{2014ExA....38..213K} on Mt.~Wendelstein in the Bavarian Alps. An independent transit detection from a ground-based facility serves not only for period confirmation and estimation of its uncertainty, but as evidence for the planetary nature of the transit from a common eclipse depth at different wavelengths. Multi-band transit photometry can be used to characterize the planet's atmosphere or rule out false positive detections \citep{2010A&A...510A.107M,2012MNRAS.422.3099S,2013MNRAS.436....2M,2016MNRAS.456..990C}. The limb darkening coefficients differ across photometric bands and can be used to differentiate between planetary signals and those of shallow-eclipse EBs. K2-95 was followed up in the i'-band on UT April 16 2016 during suboptimal weather with seeing between 1$''$ and 3$''$ and cirrus activity which led to aborting the observations after about three hours, or around mid-transit. However, due to the relative isolation of the target and reference stars on the CCD, the data was still salvageable and we could identify the transit after binning the data in 30\,min intervals. The light curve, seen in Figure \ref{fig:WST-transit}, shows the expected transit depth of 0.7\% and agrees very well with the overlaid best-fitting transit model from the \emph{K2} data, adjusted for the respective i'-band limb darkening coefficients. This light curve is already time-corrected and indicates a slight shift in phase. This implies that our initial period estimate may have been off by a few seconds per cycle, an effect seen in follow-up of previous \emph{K2} planet discoveries \citep[see][]{2016arXiv160301934B}, but it's still inside of the period uncertainty (see also \S~\ref{sec:planet-param}) of $\approx 60$\,sec. Following up transiting planets over larger baselines and therefore improving period accuracy is a valuable step in preserving the ephemeris for future studies.

\subsection{IRTF/SpeX}
\begin{figure*}
\centering
\includegraphics[width=1\linewidth]{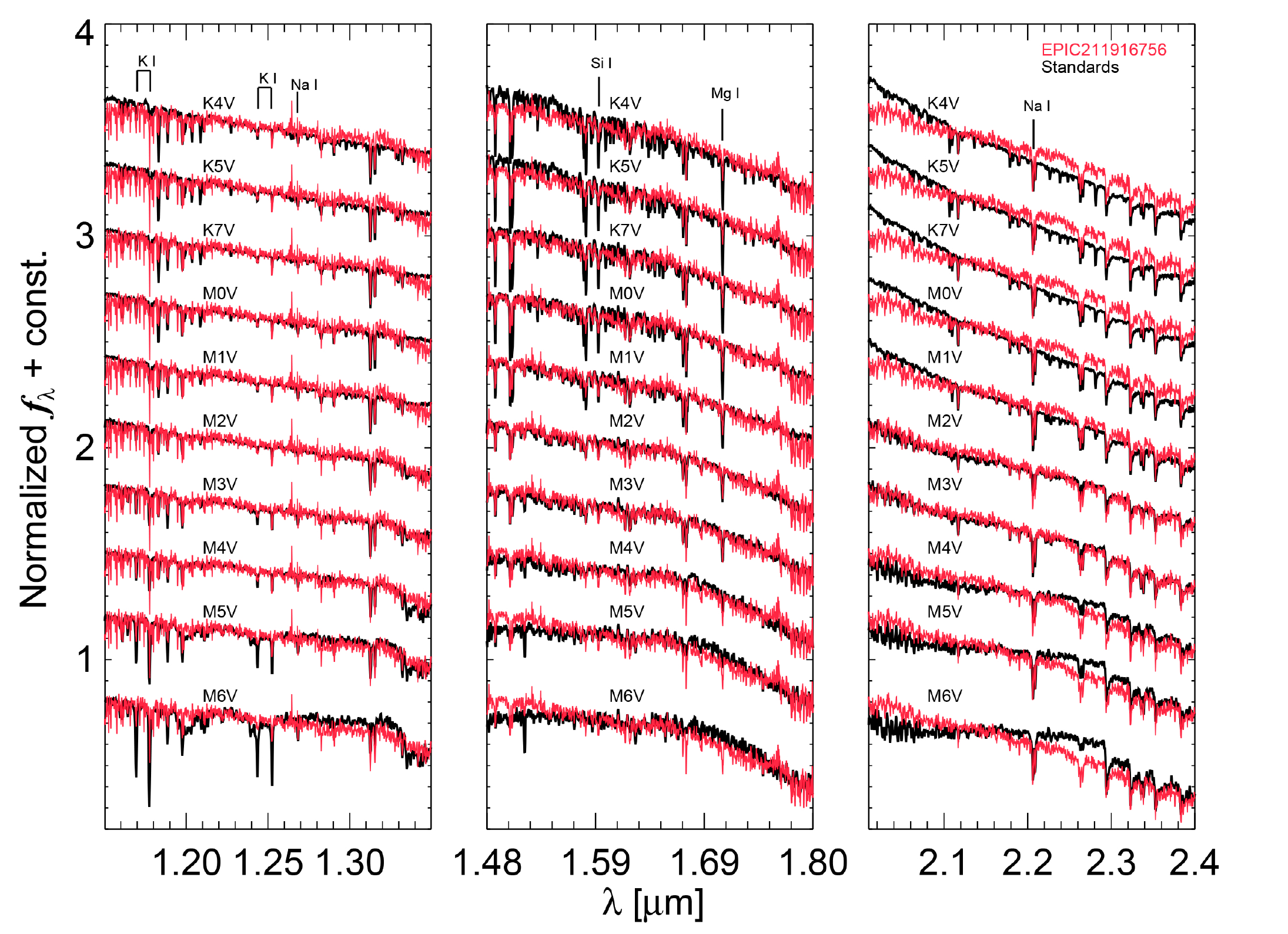}
\caption{JHK-band IRTF/SpeX spectra of K2-95, compared to K4V-M6V standard spectra from the IRTF spectral library. Every spectrum is normalized to the continuum. 
The target is a best visual match for types M2V and M3V in all three bands, which is very clear in the K-band. This is consistent with both our SED fitting results and the 
spectral typing using spectroscopic indices.}
\label{fig:SpeX}
\end{figure*}

We observed our target with the near-infrared cross-dispersed spectrograph \citep[SpeX,][] {2003PASP..115..362R} on the 
3.0\,m NASA Infrared Telescope Facility on Maunakea. While \textit{K2} targets are already pre-characterized with broadband photometry, spectral typing is essential for more accurate stellar properties. K2-95 was observed on UT December 09 2015  under excellent conditions with a clear sky and an average seeing of 0.5$''$. We used the instrument's short cross dispersed mode (SXD) with the 0.3 x 15$''$ slit which provides a wavelength range of 0.68-2.5$\,\mu m$ and a resolution of R $\approx$ 2000. The target was placed at two locations along the slit and was observed in an ABBA pattern with 16$\times$185s integrations for a total integration time of 2960s. For telluric correction and wavelength calibration, we observed an A0 standard star plus arc and flat lamp exposures right after the target.
We reduced the data with the SpeXTool package \citep{2003PASP..115..389V,2004PASP..116..362C} which performs flat fielding, sky subtraction, bad pixel removal and subsequently spectral extraction and combination, telluric correction, wavelength+flux calibration and order merging. We achieved a median signal-to-noise ratio (SNR) of 70 per resolution element in the J- ($1.25\, \mu m$), 80 in the H- ($1.6\, \mu m$) and 60 in the K-band ($2.2\, \mu m$). We compare the JHK-band spectra to late-type standards from the IRTF Spectral Library \citep{2009ApJS..185..289R}, seen in Figure \ref{fig:SpeX}. The best visual match for K2-95 lies between M2 and M3 standards across all infrared bands.

\subsection{Keck/HIRES}
We obtained a high-resolution optical spectrum of K2-95 using the HIRES echelle spectrometer on the 10m Keck I telescope \citep{1994SPIE.2198..362V} on UT December 23 2015. High-resolution spectroscopy can be used to rule out false-positive detection scenarios such as eclipsing binaries by searching for secondary line features that are created by a possible companion star.
Our observation followed the procedures of the California Planet Search 
\citep[CPS,][]{2010ApJ...721.1467H}. We used the ''C2'' decker, providing a spectral resolution of R = 55000, and subtracted the sky from the stellar spectrum. We utilized the HIRES exposure meter to automatically terminate the exposure when SNR = 32 per pixel was achieved. The HIRES spectrum was reduced using standard CPS procedures and cover $\sim$3600 -- 8000~\AA. Two additional spectra were obtained on UT December 24 and 29 using a redder setting of HIRES at R=48,000; these data are described in Pepper et al. (2016, in prep.).

\subsection{Keck/NIRC2}
We obtained high resolution NIR images of K2-95 using NIRC2 on the 10m Keck II telescope using the target as a natural guide star to drive the AO system. High-resolution imaging is a useful tool for constraining the probability of a blended background star. We observed the target on UT January 16 2016 in the $K$-band, following a multi-point dither pattern with integration times short enough to avoid saturation. We used the dithered images to subtract the sky background and remove dark current, then aligned, flat-fielded, and stacked the individual images. The star appears single and has no close companions within several arcseconds. To estimate the sensitivity of the NIRC2 observations, we injected fake sources with SNR = 5 into the combined image at separations that are integral multiples of the star's FWHM. We show our final image and the 5$\sigma$ sensitivity curve in the left panel of Figure \ref{fig:imaging}.

\subsection{Gemini-N/DSSI}
\begin{figure*}
\centering
\includegraphics[width=1\linewidth]{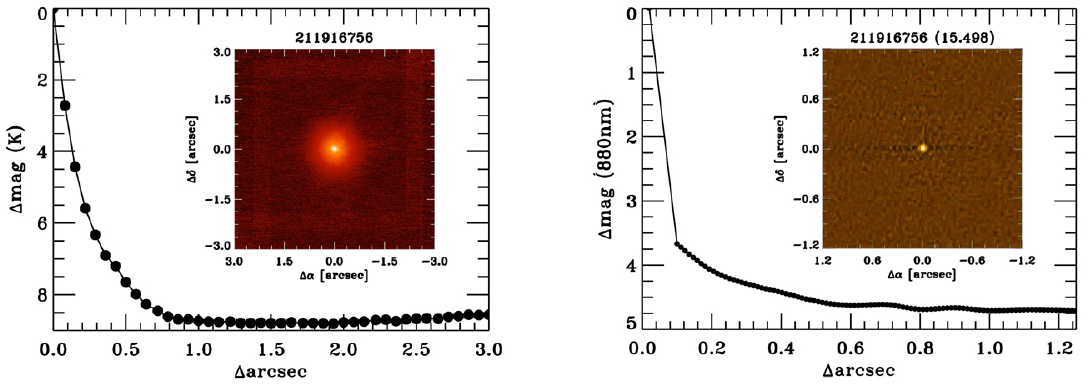}
\caption{Results from high-resolution imaging of K2-95. Left: Keck/NIRC2 $K$-band image and contrast curve. Right: Gemini-N/DSSI 880\,nm reconstructed image and 
contrast curve. The star appears single in both images and the sensitivity curves rule out the majority of close companions or background stars that would contribute significant 
flux to the transit light curve.}
\label{fig:imaging}
\end{figure*}

We also obtained speckle imaging of K2-95 in two narrow band filters centered at 880 nm and 692 nm using the DSSI camera \citep{2009AJ....137.5057H} on the 8m 
Gemini North telescope on UT January 16 2016. We followed a standard observing procedure where the star was centered in the field, guiding was established, and many images were taken 
using 60 ms exposures. The data were reduced and combined into a final reconstructed image using the techniques described in \citet{2011AJ....141...45H} and \citet{2012PASP..124.1124H}. These 
procedures perform automatic model fits (single, double, triple) and provide estimates of the magnitude difference and separation for multiple systems. K2-95
was found to be a single star. We measured the background sensitivity of the reconstructed DSSI image, using a series of concentric annuli centered on the target. The innermost
annulus is at the telescope diffraction limit where our sensitivity is zero. The sensitivities in the subsequent annuli are interpolated using a cubic spline to produce a 
smooth sensitivity curve. The 880\,nm reconstructed DSSI image and sensitivity curve are shown in the right panel of Figure \ref{fig:imaging}.

\subsection{Archival imaging}
\begin{figure*}[hbt]
\centering
\includegraphics[width=\linewidth]{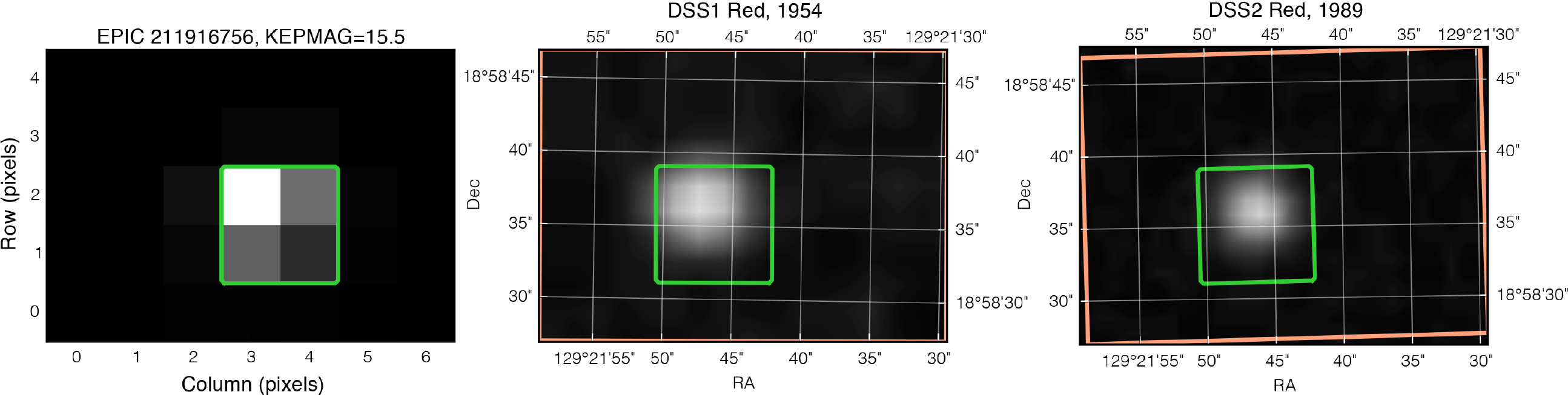}
\caption{\emph{K2} photometry with the pixels used for the light curve creation (left). DSS plates observed in red in 1954 (middle) and 1989 (right). The square shows the dimensions and location of the aperture that was used for the candidate's photometry. Over these past 35 years, K2-95 moved about 1.4\,arcsec, which is noticeable in comparison of both images.}
\label{fig:AI}
\end{figure*}

Data taken from photographic plates, now digitally scanned and available online\footnote{http://irsa.ipac.caltech.edu/applications/finderchart/}, cover several decades of astrometry. Our target was first observed in 1954 by the Digital Sky 
Survey (DSS) in the red and blue channels with an additional epoch from 1989 and 1990, respectively. We show the DSS-red plates from 1954 and 1989 in Figure \ref{fig:AI}. The images are centered on the epoch 2015 coordinates of the target in the EPIC database (08:37:27.059, +18:58:36.07) and the K2 aperture is overlaid as a green square. 
The target's proper motion of 1.4\,arcsec over the course of 35 years results in a visible shift in position, seen in comparison of the middle and left panels in Figure \ref{fig:AI}.
There is no indication for a background star at the 2015 epoch position, based on the archival data. If there is a star still hidden in the background it must be quite faint in which case it would not significantly dilute the transit signal. 

\section{HOST STAR CHARACTERIZATION}
Validation of the transiting planet candidate and constraints on its physical parameters require detailed characterization of the host star's properties. 
We used several approaches to estimate the fundamental parameters of K2-95, including medium-resolution spectroscopy, multi-band photometry and kinematics. We
also place further constraints on close bound companions and background stars from our high-resolution spectroscopy and imaging. The results of these data are used to perform a false positive probability analysis of the planet 
candidate and estimate its properties. The final stellar properties are shown in Table~\ref{tab:properties}.
\begin{table}[hbt]
\centering
\begin{tabular}{ccc}
\hline
Parameter & K2-95 & Reference  \\\hline
Epoch & J2000 & 1\\
RA & 08:37:27.059 & 1 \\
DEC & +18:58:36.07 & 1 \\
$\mu_{\alpha}$ &  $\mathrm{-36.7\pm 3.0\,mas~yr^{-1}}$      & 2 \\
$\mu_{\delta}$ &    $\mathrm{-15.1\pm 3.0\,mas~yr^{-1}}$     & 2  \\
RV &$\mathrm{35.2\pm 0.2\,km~s^{-1}}$ & 3 \\\hline
K$_p$ & 15.498\,mag& 1\\
g' & $17.779\pm 0.00240$\,mag& 4\\
r' & $16.596\pm 0.00110$\,mag& 4\\
i' & $15.369\pm 0.00079$\,mag& 4\\
z' & $14.789\pm 0.00096$\,mag& 4\\
y' & $14.529\pm 0.00220$\,mag& 4\\
J & $13.312\pm 0.01700$\,mag& 5\\
H & $12.738\pm 0.02300$\,mag& 5\\
K & $12.474\pm 0.01900$\,mag& 5\\\hline
Spectral Type & $\mathrm{M3.0\pm0.5}$  & 6\\
T$\mathrm{_{eff}}$ & $\mathrm{3471 \pm 124\,K}$ & 6, 8\\
T$\mathrm{_{eff}}$ & $3384\pm 100$\,K & 7\\
d & $\mathrm{171\pm 15}$\,pc & 7\\
d & $\mathrm{172\pm 14 }$\,pc & 3\\
$\mathrm{[Fe/H] }$& $0.11\pm 0.17$  & 6, 8\\
Radius & $\mathrm{0.402 \pm 0.050 \,R_\odot}$ & 6, 8\\
Radius & $\mathrm{0.381\pm 0.070\,R_\odot}$ & 7\\
Luminosity & $\mathrm{0.021 \pm 0.008 \,L_\odot}$ & 6, 8\\
Mass &  $\mathrm{0.361 \pm 0.069~M_{\odot}}$ & 6\\
Density &  $\mathrm{7.81 \pm 1.90 g~cm^{-3}}$& 6\\
\hline
\end{tabular} 
\caption{Stellar parameters for K2-95. References are: 1 - EPIC Catalogue; 2 - \citet{2007AJ....134.2340K}; 3 - this work; 4 - Pan-STARRS1 3$\pi$ catalog (version PV3); 5 - 2MASS catalog; 6 - this work, using \citep{2015arXiv151200483M}; 7 - this work, using SED fitting from \citet{2016A&A...587A..49O}; 8 - this work, using \citet{2015ApJ...800...85N}}
\label{tab:properties}
\end{table}

\subsection{Medium-resolution spectroscopy}
 
We apply the  index based methods of \citet{2013AJ....145...52M, 2013ApJ...779..188M, 2015ApJ...804...64M} and equivalent width (EW) based methods of
\citet{2014AJ....147...20N, 2015ApJ...800...85N} to our SpeX spectrum in order to estimate the metallicity, temperature, radius, and luminosity of K2-95.
These approaches are empirically calibrated by using wide M dwarf binary companions and nearby bright M dwarf standards with interferometrically 
measured radii. Our SpeX spectrum, shown in Figure~\ref{fig:SpeX}, suffers from poor telluric correction in the $J$ and $H$-bands. These 
residuals result from the long exposure time of the target which led to a large time baseline (nearly 1 hour) and non-ideal airmass difference ($>$0.1) between the target and A0 
calibrator. To avoid the systematic effects introduced when using the index based methods of \citet{2013ApJ...779..188M} in regions of poor telluric correction 
\citep{2013AJ....145...52M, 2015ApJ...800...85N} we use only their $K$-band relations. Prior to any analyses, the 
spectrum was shifted by its radial velocity estimated via cross-correlation with an M dwarf standard. 

To estimate the star's metallicity, we use IDL software provided by A. Mann and E. Newton\footnote{https://github.com/awmann/metal, https://
github.com/ernewton/nirew}.  Using the \citet{2013AJ....145...52M} $K$-band index relations, we estimate a metallicity $\mathrm{[Fe/H]= 0.09 \pm 0.09 \,dex}$. The $K$-band EW based methods of \citet{2014AJ....147...20N} provide $\mathrm{[Fe/H]= 0.12 \pm 0.14 \,dex}$. The uncertainties were estimated using Monte Carlo sampling. 
These estimates are consistent with each other and also with the 
metallicity of Praesepe, $\mathrm{[Fe/H]= 0.12 \pm 0.04 \,dex}$ \citep{2013ApJ...775...58B}. 

We estimate the effective temperature using the $K$-band index relations of \citet{2013ApJ...779..188M} and the $H$-band EW-based relations of \citet{2015ApJ...800...85N} using IDL software 
provided by A. Mann and E. Newton\footnote{https://github.com/awmann/Teff\_rad\_mass\_lum,  https://github.com/ernewton/nirew}. The $K$-band relations provide  $\mathrm{T_{eff} = 3460 \pm 73\,K}$ where the 
adopted uncertainty is the scatter in the polynomial relation. The $H$-band relations yield $\mathrm{T_{eff} = 3481 \pm 100\,K}$. The uncertainty was estimated using Monte Carlo sampling of the 
measurement error in the spectrum. These consistent effective temperatures are used to estimate the radius and luminosity of the star using the aforementioned empirical calibrations. Following the 
\citet{2013ApJ...779..188M} relations, we estimate $\mathrm{R_{*}=0.393 \pm 0.036~R_{\odot}}$ and $\mathrm{L_{*}=0.017 \pm 0.006~L_{\odot}}$.  The \citet{2015ApJ...800...85N} relations provide 
$\mathrm{R_{*}=0.411 \pm 0.034~R_{\odot}}$ and $\mathrm{L_{*}=0.024 \pm 0.006~L_{\odot}}$. These fundamental parameters, estimated by using different methods, are consistent at the $<1\sigma$
level. We adopt the means of these estimates for further analyses and calculate conservative uncertainties by adding the individual errors in quadrature. The final values are provided in Table~\ref{tab:properties}.  The methods of \citet{2013ApJ...779..188M} also provide 
estimates of the star's mass and density, $\mathrm{M_{*}=0.361 \pm 0.069~M_{\odot}}$ and $\mathrm{\rho_* = 7.81 \pm 1.90~g~cm^{-3}}$, respectively. We further used the H20\_K2 index 
\citep{2012ApJ...748...93R} to estimate the spectral type of the star. We find K2-95's type to be M3.0 $\pm$ 0.5, consistent with visual comparisons to standard stars and our spectroscopic temperature estimates.

\subsection{SED fitting}
We utilize the SED fitting code from \citet{2016A&A...587A..49O} as an additional layer of our stellar type characterization. In contrast to spectroscopy, this 
approach relies on broad-band photometry. We extract the Pan-STARRS1 3$\pi$ data (version PV3) for this star and cross-match its coordinates with the 2MASS catalog. For the synthetic stellar SED catalog, we use the newest version of the PARSEC isochrones package \citep{2012MNRAS.427..127B} which includes improvements for low-mass stars that were calibrated for Praesepe \citep{2014MNRAS.444.2525C}. %As in \citet{2016A&A...587A..49O}, we combine and interpolate those isochrones with the Dartmouth \citep{2008ApJS..178...89D}, Bt-Dusty \citep{2012RSPTA.370.2765A} and \citet{2015A&A...577A..42B} isochrones. 
The age of the cluster is known \citep{2015ApJ...807...24B}, therefore we restrict the synthetic model population to 800\,Myr and Praesepe's metallicity of ($\mathrm{[Fe/H]= 0.12 \,dex}$). Since the isochrone models are for nonrotating stars, we furthermore include a second set of isochrones at 650\,Myr.
We create a 10th order polynomial to interpolate between the distance-dependent extinction values given in the 3D dust map from \citet{2015ApJ...810...25G}\footnote{http://
argonaut.rc.fas.harvard.edu/} and iteratively fit distance and extinction until both converge.
We find that the final photometric fits for temperature and radius, $\mathrm{T_{eff} = 3386\pm 100\,K}$ and $\mathrm{R_* = 0.43 \pm 0.070 R_{\odot}}$, agree very well with the spectroscopic results and the extinction is negligible with $\mathrm{E(B-V)=0.0016}$. The better fit was for the 650\,Myr model with a marginally better $\chi^2$ of 7.83 against 7.97. We also estimate a distance of 171\,$\pm$15\,pc which is consistent with a Praesepe cluster membership and the derived distance of 172\,$\pm$14\,pc based on kinematic distance and K-band magnitude.

\subsection{High-Resolution Spectroscopy}

We use the methodology and algorithm of \citet{2015AJ....149...18K} to search for blended background stars or close spectroscopic binary companions 
in our HIRES spectrum. The secondary line analysis compares the observed spectrum to a suite of about 600 well characterized, slowly rotating HIRES spectra of FGKM stars from 
the California Planet Search and attempts to identify residuals consistent with a fainter secondary star. For faint, late-type stars like K2-95, this method is sensitive to spectroscopic 
companions projected within one half the HIRES slit width (0.4$^{\prime\prime}$), with approximate $V$-band fluxes as small as 3\% of
the primary flux and $\mathrm{\Delta RV > 10\, km~s^{-1}}$. This sensitivity range complements our high-resolution imaging. The algorithm also measures the 
barycentric corrected primary RV using telluric lines. The analysis revealed no secondary lines within the above sensitivity limits. 
Using the color-temperature conversions of \citet{2013ApJS..208....9P}, we estimate that the \citet{2015AJ....149...18K} analysis of our HIRES spectrum rules out a large range of close companions on circular orbits 
down to $\sim$M5.5 types on $\sim$75 day or shorter orbits.  Additionally, we measure $\mathrm{RV = 35.2 \pm 0.2~km~s^{-1}}$, consistent with other Praesepe members. The combined RV constraints from our multi-epoch HIRES observations are described further in \S~\ref{sec:fpp}.

\begin{figure}
\centering
\includegraphics[width=0.9\linewidth]{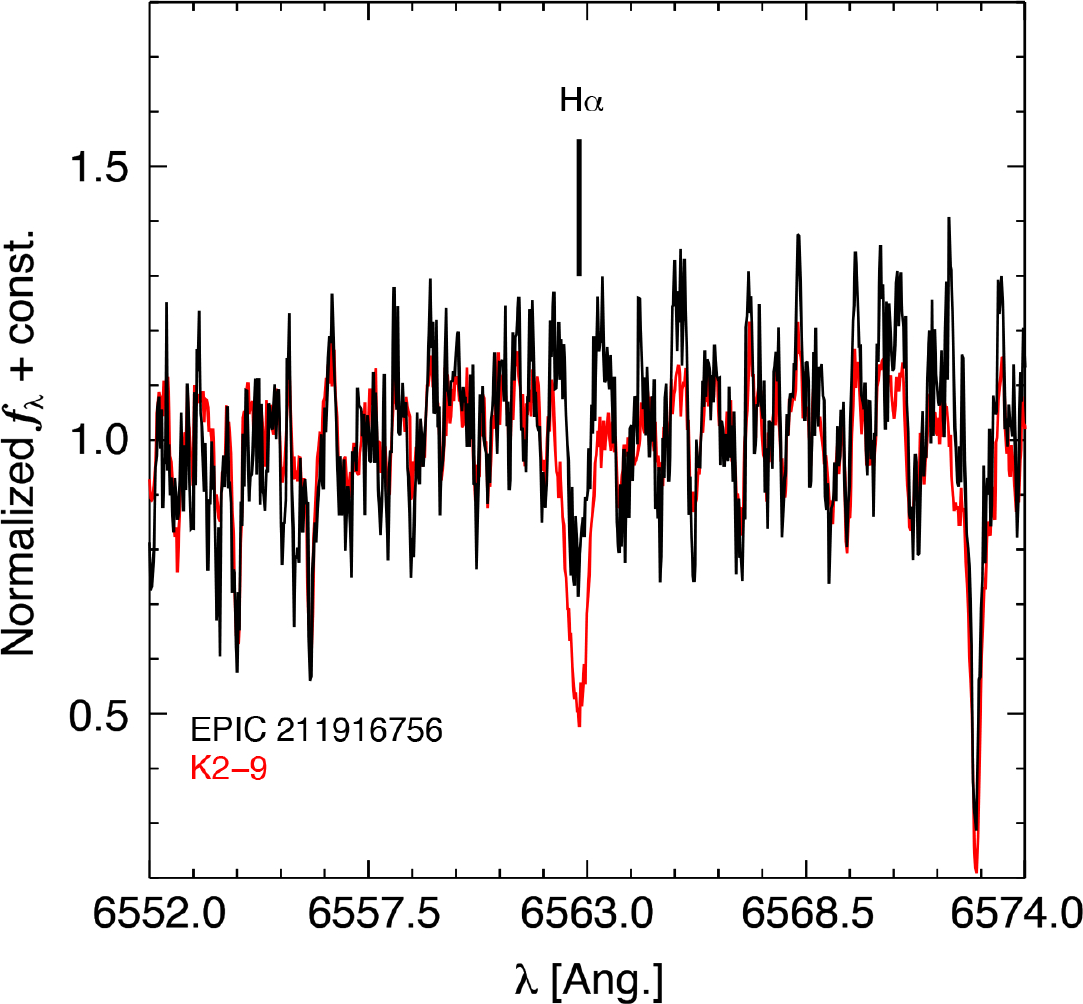}
\caption{HIRES spectrum of K2-95 (black) centered on the H$\alpha$ line compared to the known M dwarf planet host K2-9 (red). The weak activity is consistent with the lower end of the the distribution for similar spectral type stars in Praesepe and the star's slow rotation.}
\label{fig:halpha}
\end{figure}

We also use the HIRES spectrum to investigate H$\alpha$ emission at 6563~\AA. H$\alpha$ emission is a magnetic activity indicator in low-mass stars and can be used to place 
coarse constraints on a star's age \citep{2008AJ....135..785W}. \citet{2006AJ....132.1517K} and \citet{2014ApJ...795..161D} present H$\alpha$ measurements for low-mass Praesepe 
members, including K2-95. They find that M3 type stars in Praesepe exhibit a wide range of emission levels, with equivalent widths (EWs) spanning approximately
 0 to -8~\AA~(where negative EWs represent emission). K2-95 is on the low end of the emission distribution for 
 stars of similar spectral type in their studies, with only a hint of weak 
emission. We show a portion of our HIRES spectrum surrounding H$\alpha$ in Figure~\ref{fig:halpha} compared to a field age planet host with similar spectral type, K2-9 \citep{2015ApJ...809...25M,2016ApJ...818...87S}. The H$\alpha$ line morphology of K2-95 is different from the weak absorption observed in the older star K2-9, it exhibits narrow emission peaks in the line wings. This profile is consistent with model predictions for weakly active low-mass dwarfs \citep{1979ApJ...234..579C} and similar to H$\alpha$ profiles observed for the slowest rotating M dwarfs in the younger Pleiades cluster \citep[P$\sim$15 days,][]{2016arXiv160600057S}. We conservatively estimate $\mathrm{EW_{H\alpha} = -0.1 \pm 0.1}$ which is consistent with previous EW measurements and broadly consistent with expectations for an M3 dwarf in Praesepe. 

We further cross-correlated our HIRES spectrum with a slowly rotating, rotationally broadened M dwarf standard to place constraints on the projected rotational velocity $v$\,sin$i$. This analysis revealed that the star has a low rotational velocity with the best-match broadened spectrum having $v$\,sin$i$ $<$ 3 km s$^{-1}$. This low $v$\,sin$i$ and the long rotation period ($\sim$24 days) estimated from de-trended \emph{K2} photometry are consistent with the slowest rotating Praesepe M dwarfs presented in \citet{2014ApJ...795..161D}. Both indications of slow rotation are also consistent with the low level of magnetic activity inferred from the H$\alpha$ line. The slow rotation of this intermediate age M dwarf is remarkable when considering its close in planet (see \S\ref{validation}) and may indicate differences in angular momentum evolution due to initial conditions, the primordial disk, planet formation, or planet migration. In contrast, the very similar Hyades M dwarf planetary system K2-25 is among the fastest rotating M dwarfs in that cluster with a period of $\sim$1.9 days \citep{2014ApJ...795..161D,2015arXiv151200483M, 2016AJ....151..112D}.

\subsection{High-Resolution Imaging}
Using the Gemini/DSSI speckle results, we can constrain the contamination from nearby sources. The DSSI data in the 880\,nm band provide the best constraints to bound and background companions at very close separations. At a separation of 0.1$''$, our sensitivity to companions is $\mathrm{\Delta mag (880\,nm) \approx}$ 3.5\,mag.% which, using the extended online tables given by \citet{2013ApJS..208....9P}, corresponds to an M6.5 dwarf as the upper limit. 

Our Keck/NIRC2 AO imaging provides deeper constraints on close background and bound companions at larger separations. At separations of 0.2$^{\prime\prime}$ and 0.5$^{\prime\prime}$, we estimate sensitivity to companions with $\Delta$K $\approx$ 5 mag and $\Delta$K $\approx$ 8 mag, respectively. This effectively rules out all background sources within these separations that could contribute significant flux to the light curve. We use the relations of \citet{2013ApJS..208....9P} to estimate that our combined Keck and Gemini imaging rule out all bound companions at the same distance down to the hydrogen burning limit at separations of 0.1$^{\prime\prime}$ (17 AU) and well into the brown dwarf regime at $\gtrsim$ 0.5$^{\prime\prime}$ (86 AU). We use both our Keck/NIRC2 and Gemini/DSSI contrast curves as constraints in the false positive probability analysis.

\subsection{Cluster Membership, Kinematics, and Age}\label{sec:kin}

K2-95 was first identified as a candidate member of Praesepe by \citet{1994PASP..106..817W} and was subsequently included in the proposed member 
lists of several works including \citet{1995A&AS..109...29H} and \citet{2002AJ....124.1570A}. \citet{2007AJ....134.2340K} combined photometry, astrometry, and the kinematics of well defined cluster members in a maximum likelihood analysis to estimate that K2-95 has a 
$>$99\% probability of cluster membership. To further investigate its Praesepe membership, we use the star's partial kinematics and the methods described in \citet{2009AJ....137.3632L}
to estimate a kinematic distance ($\mathrm{d_{kin}}$) and predicted radial velocity (RV$_p$). In the analysis we adopt the UVW Galactic velocities 
of Praesepe from \citet{2009A&A...497..209V}
and estimate errors using Monte Carlo sampling. 
We find $\mathrm{d_{kin} = 172 \pm 14~pc}$, consistent with our SED-based estimate of the star's distance and the average cluster distance, and 
$\mathrm{RV_p = 34.1 \pm 0.9~km s^{-1}}$, consistent with our measured RV from Keck/HIRES spectroscopy. The consistency of these predictions and measurements, along with 
the spectroscopic indications of activity in our HIRES data, confirms the 
membership of K2-95 in the low-mass population of Praesepe which places a conservative constraint on its age of 600-800 Myr. We also use the kinematic distance 
and $K$-band magnitude of the star to determine its luminosity using the conversions of \citet{2013ApJS..208....9P}. We estimate $\mathrm{L_{*}=0.021 \pm 0.003~L_{\odot}}$. 
At the age of Praesepe, an M3 dwarf is expected to be on the main sequence and has stopped radial contraction. We can therefore combine our 
measured effective temperature and luminosity through the Stefan-Boltzman law to estimate the star's radius, $\mathrm{R_{*}=0.40 \pm 0.01~R_{\odot}}$. These alternate estimates of the star's
fundamental parameters are consistent with those from our SpeX spectroscopy and SED fitting.

\section{PLANET VALIDATION}\label{validation}
\subsection{False positive probability}\label{sec:fpp}

For a transiting planet-signal, there are five common sources of false-positive identification or transit mischaracterization, most of which are created by eclipsing binaries (EB's):
\begin{itemize}
\item Background star
\item Blended EB system
\item Unblended EB system
\item Double-period EB system
\item Hierarchical EB companion
\end{itemize}
Our collected data in form of photometry, spectroscopy and high-resolution imaging can be used to place a number of constraints on the data to limit or even completely rule out all of the above scenarios. In the \emph{K2} data itself, we detected no secondary eclipse that would be indicative of an EB. Based on archival and high-resolution imaging and high-resolution spectra, a background source is strongly constrained to less than 3\% of flux dilution and can be ruled out completely for a separation of more than 0.2\,arcsec. This makes any kind of background blend or triple system highly improbable. In case it did exist, it would not impact the planet parameters significantly. 

For a more quantitative assessment, we utilize the false positive probability (FPP) calculator \textit{vespa} \citep{2012ApJ...761....6M,2015ascl.soft03011M} which is open source and freely available online\footnote{https://github.com/timothydmorton/vespa}. This program compares the light curve to transit shapes created by false-positive sources and combines this with priors about stellar population, multiplicity frequencies and the planet occurrence rate for the corresponding fitted parameters. We supply the algorithm with all of our determined constraints, including stellar photometry from 2MASS and WISE, contrast curves from high-angular resolution imaging and the light curve from \emph{K2}. Furthermore, we also extract the photometric light curve from \citet{2014PASP..126..948V}, remove the periodic modulations, recover the signal with the Pan-Planets signal detection pipeline \citep{2016A&A...587A..49O} and then perform the same analysis. This way, we end up with an independent confirmation based on a different data reduction and signal detection routine.  Based on all of the above constraints, the results from \textit{vespa} rule out all false positive scenarios to a FPP of less than 0.02\% for both analyses. While \textit{vespa} does not fit blended planetary systems, there are strong constraints on this scenario based on high-resolution imaging and the upper limit of 3\% in flux dilution for background sources which makes this scenario highly unlikely. As an additional layer of security, we furthermore obtained three RV points based on high-resolution spectroscopy in order to constrain any EB or double-period EB scenario.

\subsubsection{Unblended EB system}
The unblended EB scenario consists of very shallow eclipses of both stars which may emulate a planet's transit light curve. There are many constraints to this scenario in the case of K2-95: the signal of a secondary eclipse is absent in the light curve data and the high-resolution spectroscopy excludes the presence of a second star down to 10\,km\,s$^{-1}$ and 3\% flux. Based on both our own observation with HIRES and the two additional data points from Pepper et al. (2016, in prep.), we cover a time baseline of ~6 days that we use to construct a 5$\sigma$ upper limit for the maximum RV amplitude that could still fit to the data and is shown in Figure \ref{fig:RV} in the top panel. The result is an amplitude 941$\mathrm{\,m~s^{-1}}$ which equates to 5.25\,$\mathrm{M_J}$, a giant planet. These limitations mean that this signal can not be modeled as an unblended EB system.
\subsubsection{Double-period EB system}
The double-period case is different to other scenarios in that it assumes an EB system in which both partners have the same size and eclipse each other. This changes fundamental parameter such as the relative eclipse duration and impacts limits for the secondary eclipse - strong constraints make this scenario more likely. %It was the only scenario for which \textit{vespa} estimated a noteworthy probability of about 1\%. However, additional factors help in ruling out this scenario completely. 

\begin{figure}
\centering
\includegraphics[width=1\linewidth]{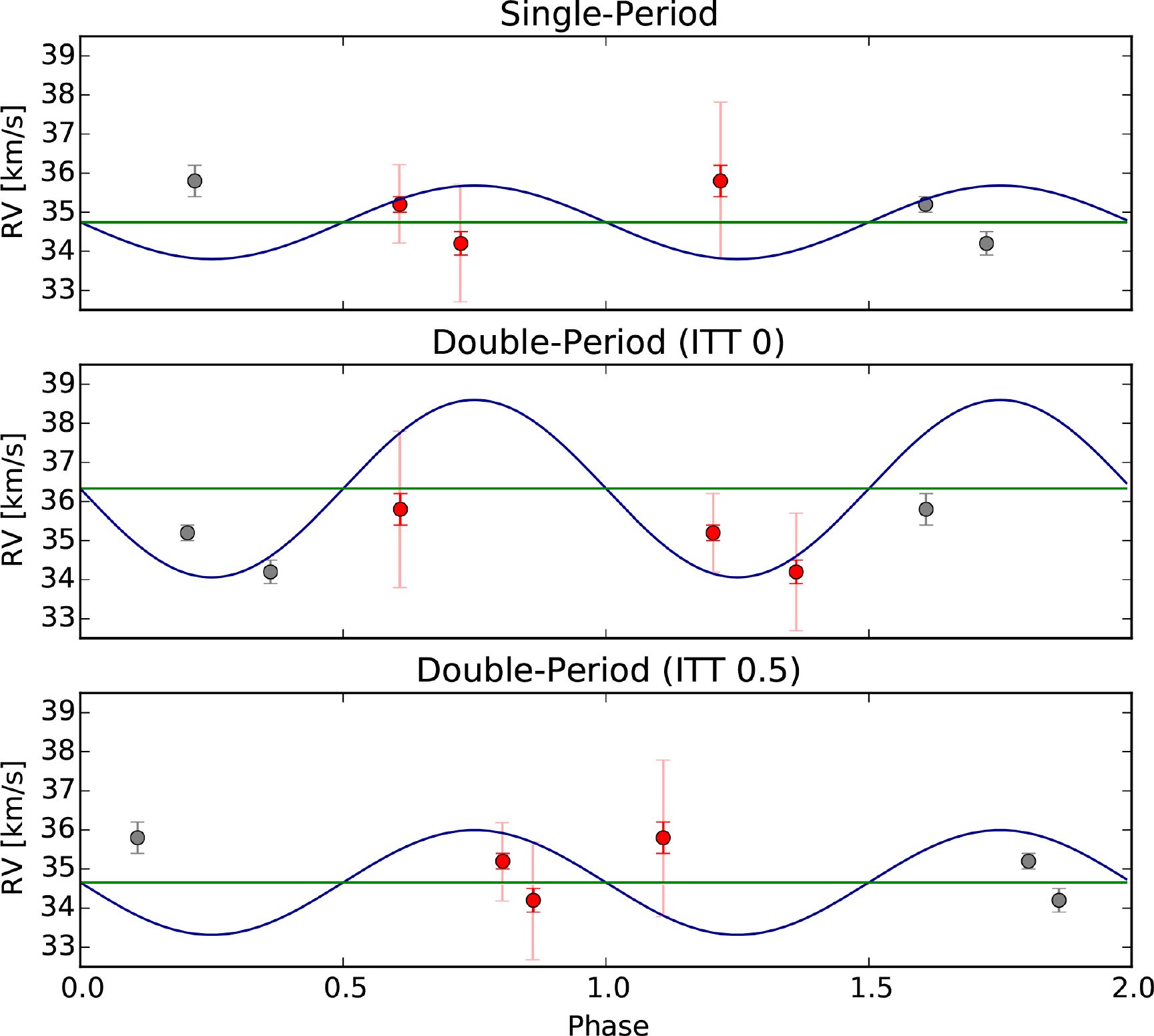}
\caption{Radial velocity for K2-95 in the single-period (top) and double-period scenario (middle+bottom), phased to the corresponding period and ITT scenario. ITT stands for the initial time phase of the first recorded eclipse, i.e. whether the primary or secondary star eclipsed first. The RV curve (blue) shows the maximum amplitude consistent with the points at 5$\sigma$. Two phases are shown for better clarity with repeated points grayed out and the error bars of the points are given in 1$\sigma$ (red) and 5$\sigma$ (light red). The green line shows the baseline fit.}
\label{fig:RV}
\end{figure}

Both partners must have similar radii in this case. 
As in the single-period EB scenario, we combine our HIRES RV measurement with the two measurements presented in Pepper et al. (2016, in prep.). % to cover a total time baseline of ~6 days. These RV measurements cover a substantial portion of the phase of the transiting planet candidate and allow us to construct an RV curve. 
We again set limits for which the RV curve is outside of 5$\sigma$ of the individual points, which is shown in Figure \ref{fig:RV} in the middle and bottom panels. Two cases have to be considered, depending on whether the initial transit time (ITT) was at phase 0 or 0.5 (ITT 0 and ITT 0.5, respectively). The subsequent limit is a RV of 2270$\mathrm{\,m~s^{-1}}$ for ITT 0 and 1343$\,\mathrm{m~s^{-1}}$ for ITT 0.5. Taking the stellar mass determined by medium-resolution spectroscopy and assuming a circular orbit, this translates to 15.46$\mathrm{\,M_J}$, a low-mass brown dwarf, or 9.14$\mathrm{\,M_J}$, a giant planet. Any stellar companion would produce a much stronger RV signal and an eclipse of the primary in front of a brown dwarf cannot create such a strong signal.

Additionally, K2-95 has a probability of more than 99\% for being a member of the Praesepe cluster, which means that the baseline of the fitted RV curve for the case of ITT 0, $\mathrm{RV = 36.3~km~s^{-1}}$, should be consistent with the cluster $\mathrm{RV_p = 34.1 \pm 0.9~km s^{-1}}$. ITT 0 is only consistent at 3$\sigma$ which further decreases the likelihood of this scenario. In contrast, the RV baseline for a single-period transiting planet scenario is very consistent with a best fit of 34.8$\,\mathrm{kms^{-1}}$. 

Therefore, in combination with all of the other constraints (e.g. AO imaging, archival optics, stellar characterization), the transit signal can not be modeled successfully with this scenario and we can rule it out.
\subsection{Planet parameters}
\label{sec:planet-param}
We analyze the light curve of K2-95 with a approach similar to the one described in more detail by \citet{2015ApJ...804...10C}\footnote{Further information about the most up- to-date method will be found in Crossfield et al. (2016), in prep.}. In brief: Relying on the \textit{emcee} package \citep{2013PASP..125..306F},  we use the open-source BATMAN light curve code  \citep{2015PASP..127.1161K} which we optimized for long-cadence data.
Utilizing the free and open-source LDTk/pyLDTk package from \citet{2015MNRAS.453.3821P}\footnote{https://github.com/hpparvi/ldtk}, we propagate our measured $\mathrm{T_{eff}}$, surface gravity, metallicity and their respective uncertainties into limb-darkening coefficients for use as priors in our fit.  The overall fitted parameters in our analysis are the candidate's orbital period P, initial transit time $T_0$,  inclination $i$, eccentricity $e$, longitude $\omega$, scaled semi-major axis $a/R_\star$ and the fractional candidate radius $R_p/R_\star$. The starting parameters for the fit are taken from our TERRA output. In the fit, we assume a linear ephemeris for the transits which should be a valid simplification since there is no evidence for any kind of TTV's in the light curve. The best-fitting properties and their uncertainties are shown in Table~\ref{tab:planet-properties}.
\begin{table}[hbt]
\centering
\begin{tabular}{llc}
\hline
Parameter & Units  & K2-95  \\\hline
$\mathrm{T_0}$&  BJD$\mathrm{_{TDB}}$ - 2454833&  $\mathrm{2338.1477^{+0.0018}_{-0.0019}}$ \\
P  & d  & $10. 13389^{+0.00068}_{-0.00077}$  \\
$i$  & deg  & $88. 77^{+0.86}_{-1.59}$  \\
$R_P/R_\star$  & \%  & $7.86^{+1. 69}_{-0.93}$  \\
$\mathrm{R_\star /a }$  & ---  & $0. 0400^{+0.0187}_{-0.0068}$  \\
$T_{14}$ & hr&  $2.84^{+0.36}_{-0.26}$\\
$T_{23}$ & hr& $2.18^{+0.26}_{-0.72}$\\
$a$  & AU  & $0. 0653^{+0.0039}_{-0.0045}$  \\
$R_P$  & $R_E$  & $3. 47^{+0.78}_{-0.53}$  \\
$R_\star$  & $R_\odot$   & $0.402^{+0. 050}_{-0.050}$  \\
$M_\star$  & $M_\odot$   & $0.361^{+0. 069}_{-0.069}$  \\
\hline
\end{tabular} 
\caption{Best-fitting properties of K2-95 and its planet based on the BATMAN code. }
\label{tab:planet-properties}
\end{table}
We estimate the planet's mass using the mass-radius relation\footnote{And their code: https://github.com/dawolfgang/MRrelation} provided by \citet{2015ApJ...806..183W} and \citet{2015arXiv150407557W}, $\mathrm{M/M_\oplus = 2.7(R/R_\oplus)^{1.3}}$ to $\mathrm{M_P = 13.71\pm 3.62~M_\oplus}$\footnote{The code cannot handle asymmetrical errors, hence we selected the larger of both uncertainties.}. However, using the relation provided by \citet{2014ApJ...783L...6W}, $\mathrm{M/M_\oplus = 2.69(R/R_\oplus)^{0.93}}$, we get $\mathrm{M_P = 8.77^{+1.88}_{-0.53}~M_\oplus}$. A third mass-radius relation, published by \citet{2016arXiv160308614C}\footnote{https://github.com/chenjj2/forecaster}\footnote{The code cannot handle asymmetrical errors, hence we used the larger of both uncertainties.}, yields $\mathrm{M_P = 8.26^{+1.77}_{-0.50}~M_\oplus}$ based on the relation $\mathrm{M/M_\oplus = (R/R_\oplus)^{1.70}}$. 
\begin{figure*}
\centering
\includegraphics[width=1\linewidth]{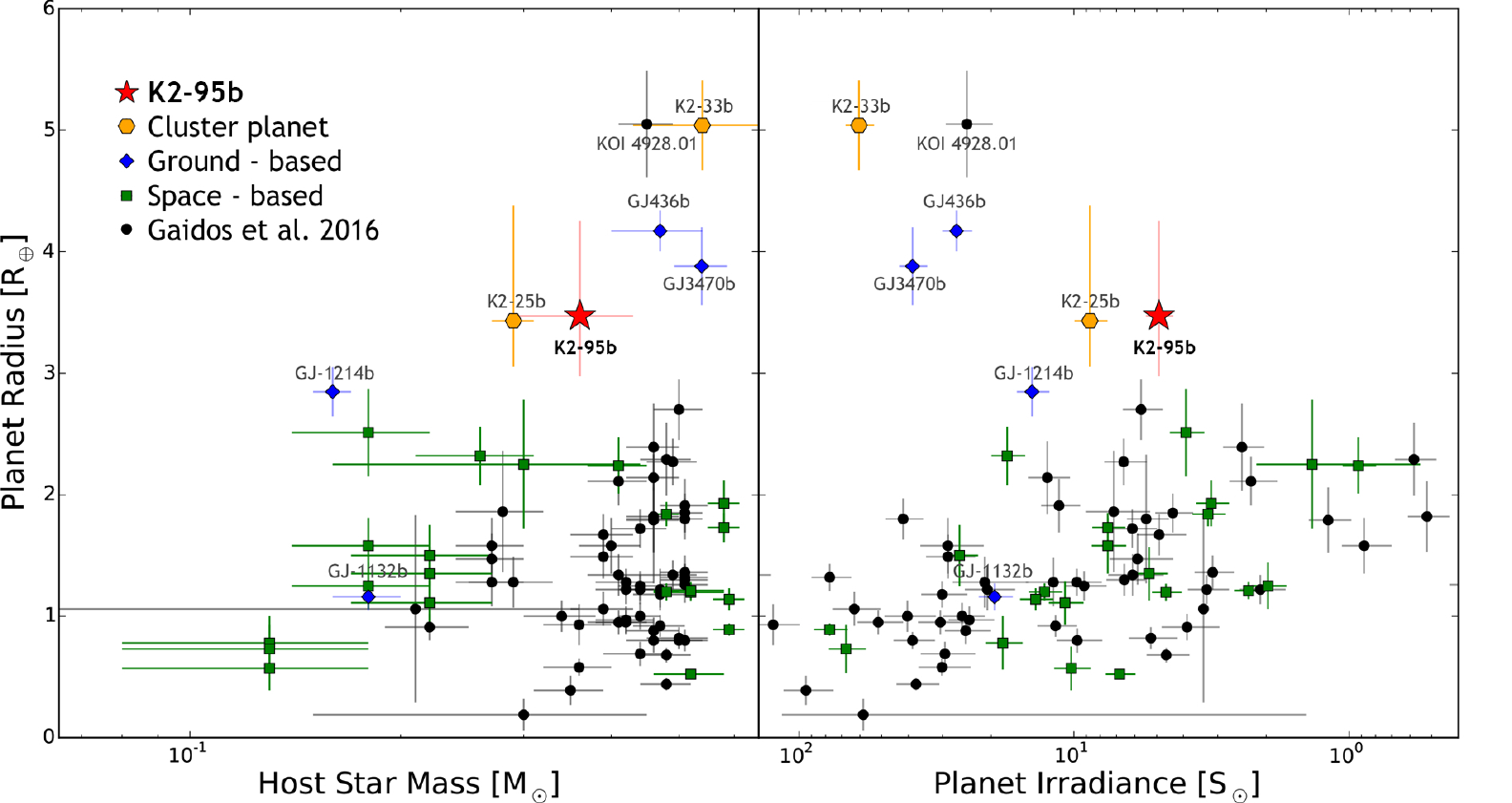}
\caption{Planet radius as a function of the host star mass (left) and received radiation (right), comparing our discovery K2-95 (red star) to planet detections in open clusters (orange hexagons), ground-based surveys (blue diamonds), space-based (Kepler+K2) surveys (green squares), and revised values for several Kepler planets from \citet{2016MNRAS.457.2877G} (black circles). Similar to \citet{2015arXiv151200483M}, only stellar radii below 0.5\,$\mathrm{R_\odot}$ and periods below 100\,d were included. Two exceptions to those criteria are RV-planet GJ 3470b and K2-33b which got added due to their similarity despite a larger host star radius. All RV detections and inflated planets are labelled.}
\label{fig:planet-stats}
\end{figure*}
The mass-radius models lead to different estimates of the planet's mass. While the results from \citet{2015ApJ...806..183W} are higher than the other two, the difference is still small enough for the masses to be marginally consistent with each other.
The absence of TTV's in the system means that the mass can not be determined through other means as of now. We estimate the RV amplitude of this planet to be 6.8\,$\pm 1.8\,\mathrm{m\,s^{-1}}$, based on the \citet{2015ApJ...806..183W} results. 
\section{DISCUSSION}

So far, only very few planets have been detected in clusters, even less with the transit method. K2-95b is only the third known planet in an open cluster that orbits around an M dwarf. Assuming a typical density of small gas planets, it probably belongs to the class of Neptune-size planets with a similar chemical composition and H/He atmospheres \citep{2014ApJS..210...20M,2014ApJ...783L...6W,2015ApJ...801...41R}.

However, it is remarkable that the occurrence rate of planets with the radius and period of both K2-25 \citep{2015arXiv151200483M,2016AJ....151..112D} and K2-95b is very low around field stars \citep{2015ApJ...807...45D,2015ApJ...814..130M}. Furthermore, the recently discovered planet K2-33b in the open cluster Upper Scorpius \citep{2016Nature...accepted, 2016arXiv160406165M} exhibits an unusually large radius as well. 
While there are four discovered systems with planet radii higher than K2-95b and K2-25b, those planets are even larger and orbit higher-mass stars. Furthermore, their received stellar flux appears to be significantly higher. 
The distribution of planetary radii and received radiation against the host star mass are shown in Figure \ref{fig:planet-stats}. We placed following restrictions: All planets in this Figure have to be confirmed and we extract the most recent planetary and stellar parameters from the NASA exoplanet archive \citep{2013PASP..125..989A}.  Furthermore, the host star radii have to be below 0.5\,$\mathrm{R_\odot}$ and the planet irradiance was calculated when missing. 

The probability of detecting two such planets in a cluster without any detections in the larger field star sample, plus another detection in a scarcely populated region of larger-radius planets, is too low to be random chance. We present three possible implications from this:
\begin{itemize}
\item The formation of short-period planets is different in clusters due to gravitational interactions during migration. An indication for this may be the higher occurrence rate of hot Jupiters in M67 measured by \citet{2016arXiv160605247B}. However, \citet{2013Natur.499...55M} found an occurrence rate similar to that of field stars for NGC 6811. As of now, there is insufficient information to confirm this theory.
\item M dwarfs remain active for several hundred Myr after their formation to a varying degree \citep{2014AJ....148...64S}. Strong UV emission in the relatively young Hyades and Praesepe cluster M dwarfs might lead to the inflation seen in Figure \ref{fig:planet-stats}. However, no emission could be detected by GALEX down to 19.9\,mag in the far UV and 20.8\,mag in the NUV \citep{2011Ap&SS.335..161B}. Young planets may also be larger due to initial heat from formation \citep{2015arXiv151200483M}.
\item It is possible that this is due to a selection bias since young stars are more active. Their variability may mask many of the small-planet transit signals, leading to a perceived imbalance. However, K2-95 is only weakly active so while a selection bias may exist, it is unlikely to be the sole reason.
\end{itemize}

Measuring the stellar UV activity and the planet's mass will allow to determine whether the reason behind the large radii is inflation due to strong UV irradiation and/or initial heat. If that were the case, they could be seen as outliers of the general planet mass-radius relation and might be similar to GJ 436b, a Neptune-sized planet first detected by RV measurements \citep{2004ApJ...617..580B} that is showing visible transits \citep{2007A&A...472L..13G} and appears to evaporate \citep{2015Natur.522..459E}. However, as it can be seen in Figure \ref{fig:planet-stats} on the right, GJ 436b receives several times of K2-95b's radiation so it is questionable whether this may apply here. Both cluster detections also orbit noticeably smaller stars than the larger Neptunian planets.

Besides this anomaly, K2-95b is also intriguing for a number of other reasons, especially for having a well-determined distance, (young) age and metallicity. Only very few planets are known around relatively young stars and new detections will contribute towards establishing a more accurate timeline of planetary development.

Assuming a circular orbit - considering the transit duration shows no indication of ellipticity a valid simplification - and using the mass-radius relation from \citet{2015ApJ...806..183W}, we calculate the radial velocity amplitude to 6.8\,$\pm 1.8\,\mathrm{m\,s^{-1}}$. While an accuracy of $\mathrm{m\,s^{-1}}$ is entirely feasible today with instruments like HIRES or HARPS, the target is too faint to realistically achieve this with today's telescopes in reasonable observing times. However, future dedicated infrared spectrographs such as IRD and HPF (\citet{2014SPIE.9147E..14K} and \citet{2012SPIE.8446E..1SM}, respectively) will allow the determination of the planet's mass. This in turn will also provide additional data for the calibration of the mass-radius relation of Neptune-sized gas planets. Next-generation large telescopes such as the E-ELT or the TMT may enable a detailed study of the planet's atmosphere. 

As an alternative to spectroscopy, multi-band photometry enables a more detailed study of the planet, even for stars that are too faint for atmosphere spectroscopy. Depending on the photometric band, the transit eclipse depth may vary due to Rayleigh scattering or varying opacities which allows to model the atmosphere \citep{2010A&A...510A.107M,2012MNRAS.422.3099S,2013MNRAS.436....2M,2016MNRAS.456..990C}. While this is possible to do with single-band photometric instruments, simultaneous multi-band capture with GROND \citep{2008PASP..120..405G} or the upcoming 3 channel imager 3KK at Mt.~Wendelstein \citep{2010SPIE.7735E..3QL} would be much more advantageous. 
\section{SUMMARY}
We report on the discovery of a Neptune-sized planet in the Beehive cluster (Praesepe) that orbits a cool dwarf star. Discussing and subsequently ruling out each possible false-positive detection scenario, we validate the planetary nature of this candidate. Using detailed follow-up, including ground-based transit recording, spectroscopy and high-resolution imaging, we characterize both the host star and its planet. We noticed a radius anomaly for this planet and the previously detected K2-25b, both planets around M dwarfs in clusters. Both of them possess radii that are in a region seemingly unpopulated by planets orbiting comparable field stars. Detailed study and future observations will reveal whether this is due to different planet formation or evolution in open clusters. 
\begin{acknowledgements}
We thank the staff of the Wendelstein observatory for technical help and strong support during the data acquisition, including observing the target for us. We especially thank Ulrich Hopp for his constructive input during and after observations.

The research of J.E.S was supported by an appointment to the NASA Postdoctoral Program at NASA Ames Research Center, administered by Universities Space Research Association through a contract with NASA.

E.A.P acknowledges support through a Hubble Fellowship.

Some of the data presented herein were obtained at the W.M. Keck Observatory (which is operated as a scientific partnership among Caltech, UC, and NASA) and at the Infrared Telescope Facility (IRTF, operated by UH under NASA contract NNH14CK55B). The authors wish to recognize and acknowledge the very significant cultural role and reverence that the summit of Maunakea has always had within the indigenous Hawaiian community. We are most fortunate to have the opportunity to conduct observations from this mountain.

Based on observations obtained at the Gemini Observatory, which is operated by the Association of Universities for Research in Astronomy, Inc., under a cooperative agreement with the NSF on behalf of the Gemini partnership: the National Science Foundation (United States), the National Research Council (Canada), CONICYT (Chile), Ministerio de Ciencia, Tecnolog\'{i}a e Innovaci\'{o}n Productiva (Argentina), and Minist\'{e}rio da Ci\^{e}ncia, Tecnologia e Inova\c{c}\~{a}o (Brazil).

\textit{Facilities}: Kepler, K2, IRTF (SpeX), Keck: I (HIRES), Keck:
II (NIRC2), Gemini-N (DSSI)
\end{acknowledgements}
\software{vespa (Morton 2012, 2015), emcee (Foreman-Mackey et al. 2013), BATMAN (Kreidberg 2015), LDTk/pyLDTk (Parviainen \& Aigrain 2015)}

%% This command is needed to show the entire author+affilation list when
%% the collaboration and author truncation commands are used.  It has to
%% go at the end of the manuscript.

%\allauthors

\end{document}